\documentclass[11pt]{article}

\usepackage[preprint]{acl}

\usepackage{times}
\usepackage{latexsym}

\usepackage[T1]{fontenc}
\usepackage[utf8]{inputenc}

\usepackage{microtype}

\usepackage{inconsolata}

\usepackage{graphicx}
\usepackage{amsmath}
\usepackage{amssymb}
\usepackage{cleveref}

\newcommand{\rui}[1]{}
\newcommand{\yc}[1]{}
\newcommand{\james}[1]{}
\newcommand{\xuan}[1]{}

\title{SidConArena: An Environment Evaluating Agents in Open-Ended, Positive-Sum Bargaining Game}

\author{
Yeqi Feng\textsuperscript{1}\thanks{Equal contribution. $^\dagger$ Corresponding authors.},
Yuxin Chen\textsuperscript{1}\footnotemark[1],
Tianxing He\textsuperscript{1 $\dagger$}\\   
\textsuperscript{1}Institute for Interdisciplinary Information Sciences, Tsinghua University\\
\texttt{\{fengyq25, chenyuxin25\}@mails.tsinghua.edu.cn}
 \\
\texttt{hetianxing@mail.tsinghua.edu.cn} \\
}

\begin{document}
\maketitle
\begin{abstract}
Evaluating LLM agents requires dynamic environments that go beyond static reasoning and zero-sum games. Real-world economic interaction is often open-ended and mixed-motive: agents must negotiate, create positive-sum surplus, compete for scarce assets, and plan under delayed returns. We introduce \textit{SidConArena}, a new benchmark framework for evaluating LLM agents in open-ended, positive-sum bargaining. SidConArena formalizes a multi-player economy as a finite-horizon partially observable stochastic game with three coupled phases: natural-language negotiation with binding trades, deterministic converter-based production, and sealed-bid auctions for long-term assets. The framework combines structured observations, phase-aware agent dispatching, a neural-symbolic action interface, and asynchronous execution, enabling free-form interaction while preserving rule-grounded evaluation. Across homogeneous and heterogeneous tournaments, stronger frontier models achieve higher economic outcomes, yet agents still misvalue resources, bargain passively, and remain limited in long-horizon investment planning.
\end{abstract}

\section{Introduction}\label{sec::introduction}

Large Language Models (LLMs) are increasingly used as autonomous agents that interact with environments and other agents. This shift calls for benchmarks that go beyond static reasoning or isolated instruction following. While existing evaluations measure important capabilities, many remain static and may suffer from contamination risks~\citep{hao2024llmreasonersnewevaluation, perlitz2024llmbenchmarksagreefixing, sainz2023nlpevaluationtroubleneed}. They therefore provide limited insight into how agents behave in dynamic, partially observable, and socially interactive settings.

Strategic games provide a natural testbed for evaluating planning and interaction. However, many existing game-based benchmarks emphasize competitive, zero-sum, or adversarial settings~\citep{paquette2019pressdiplomacymodelingmultiagent, xu2025languageagentsreinforcementlearning}. In contrast, many real-world economic interactions are mixed-motive: agents must cooperate to create surplus while competing over scarce resources. Strong performance requires not only rule following, but also local valuation, bargaining, resource allocation, and long-horizon investment planning.

To address this gap, we introduce \textit{SidConArena}, a new multi-agent benchmark framework for open-ended, positive-sum bargaining. Based on the economic structure of \textit{Sidereal Confluence}, SidConArena places asymmetric agents in a multi-player economy where they negotiate binding trades, activate production converters, and compete in sealed-bid auctions for long-term assets. The framework combines structured observations, phase-aware agent dispatching, a neural-symbolic action interface, and asynchronous execution, enabling natural-language negotiation while preserving rule-grounded evaluation. Through homogeneous self-play and heterogeneous tournaments, we show that SidConArena distinguishes model capabilities and reveals persistent limitations in valuation, bargaining, and long-horizon economic planning.
% 大语言模型在不同的任务上展现出强大的能力。现有有很多评测大语言模型能力的工作，如ABCDE，评测了大语言模型 ABCDE的能力。
% 策略游戏，尤其是桌游，对于评测大语言模型的沟通、交流、谈判、合作、博弈等能力具有价值。While 许多相关的工作聚焦于智能体在竞争环境中的能力，也有一些相关的工作考察了智能体的合作能力。如外交等等。评测智能体在复杂的异质性的 positive-sum negotiating 策略游戏中的能力很有价值。

% 我们计划构建一个名为 SidConArena 的多智能体评测框架，专门用于测试 LLM 在复杂的经济谈判中的表现。我们将完整复现桌面游戏《Sidereal Confluence》—— 一个以开放式、多方、非零和（Positive-Sum）交易为核心的复杂策略游戏。我们的目标是让不同的 LLM Agent 在这个“竞技场”中相互对抗来量化评估它们在复杂谈判、承诺遵循、长期规划和合作竞争方面的涌现能力，这将填补目前多智能体 AI 评测在复杂经济博弈领域的空白。

% We thus propose SiderConArena, a highly complex, positive-sum, open-ended-bargaining, strategic-boardgame-based environment to benchmark agents ability in cooperative compitive mixed goal or sth.?
\section{Related Work}\label{sec::related_works}

\paragraph{Strategic Game Benchmarks.}
Recent research has increasingly utilized strategic games as complex environments for evaluating the reasoning and interaction capabilities of large language models (LLMs). \citet{cipolinakun2025gamereasoningarenaframework} develop a generalized framework, \textit{Board Game Arena}, for testing LLMs across various strategic games. Other game-based benchmarks also evaluate strategic reasoning in interactive settings, such as \textit{GameBench}~\citep{costarelli2024gamebench} and \textit{AvalonBench}~\citep{light2023avalonbench}. Some works focus on \textit{Diplomacy}, where models aim to coordinate, negotiate, or eliminate others~\citep{paquette2019pressdiplomacymodelingmultiagent, bakhtin2022humanlevel, xu2025dipllmfinetuningllmstrategic}. Related studies also investigate social deduction games such as \textit{Werewolf}~\citep{xu2025languageagentsreinforcementlearning}. There are also works that utilize real-time strategy games such as \textit{StarCraft II} as benchmarks~\citep{ma2024largelanguagemodelsplay}.
\vspace{-3mm}

\paragraph{Multi-agent Benchmarks.}
There also exist benchmarks focusing on evaluating the abilities of multi-agent systems or observing their interactions. \textit{LLM-DELIBERATION} focuses specifically on negotiation~\citep{abdelnabi2024llmdeliberation}. \textit{NegotiationArena} further evaluates LLMs in scorable bargaining scenarios~\citep{bianchi2024negotiationarena}. More comprehensive testbeds such as \textit{SmartPlay}~\citep{wu2024smartplaybenchmarkllmsintelligent}, \textit{AgentVerse}~\citep{chen2023agentversefacilitatingmultiagentcollaboration}, and \textit{MultiAgentBench}~\citep{zhu2025multiagentbenchevaluatingcollaborationcompetition} provide different types of interactive tasks. General agent benchmarks such as \textit{AgentBench}~\citep{liu2023agentbench} and \textit{AgentBoard}~\citep{ma2024agentboard} also evaluate LLM agents across multi-turn decision-making environments.
\vspace{-3mm}

\paragraph{Social Simulation.}
Social simulation frameworks typically contain agents, an environment they interact with, and interfaces that mediate their interactions~\citep{gao2024xiaochong}. Existing work can be broadly divided into two categories. The first category encompasses general-purpose social simulation frameworks such as \citet{park2023generativeagentsinteractivesimulacra}, \citet{tian2025visualizedframeworkeventcooperation}, \citet{tang2024gensim}, and \citet{huang2025adasocietyadaptiveenvironmentsocial}, which aim to model general societal patterns. Recent environments such as \textit{SOTOPIA}~\citep{zhou2023sotopia} and \textit{Concordia}~\citep{vezhnevets2023generative} also study social interaction and agent-based simulation with language agents. The second category investigates specific social phenomena, such as public administration crises~\citep{xiao2023simulatingpublicadministrationcrisis}, health policy~\citep{hou2025societygenerativeagentssimulate}, political manipulations~\citep{touzel2024simulation}, and financial markets~\citep{gao2024simulatingfinancialmarketlarge}.

% 这一节具体讲一下过去相关的工作（感觉可以写 a. LLMs 玩桌游 b. LLMs 玩策略游戏 c. 多智能体相关的一些工作）（待补充）
% filepath: refined_formulation_v3.tex

\begin{figure*}
    \centering
    \includegraphics[width=1\linewidth]{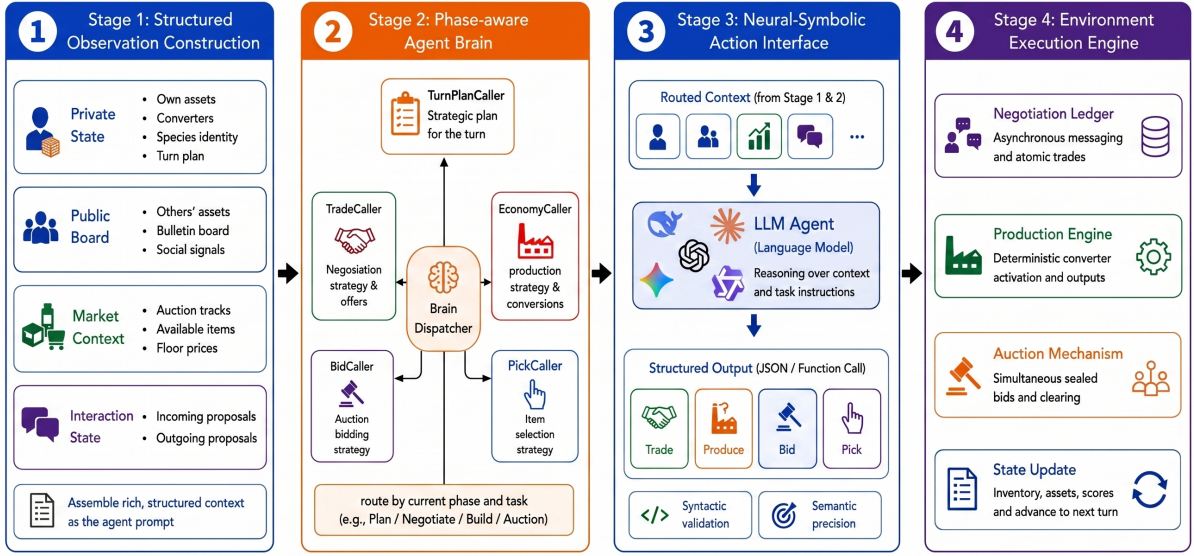}
    \caption{\textbf{Overview of the SidConArena framework.} SidConArena decomposes each agent's turn into four stages. First, the environment assembles a structured observation from private state, public board information, market context, and interaction history. Second, a phase-aware agent brain routes the task to specialized callers for turn planning, trading, production, bidding, and item selection. Third, the routed context is passed to an LLM agent through a neural-symbolic action interface, where free-form reasoning is converted into validated function-call actions. Finally, the execution engine applies the submitted actions through the negotiation ledger, production engine, auction mechanism, and state update module, advancing the game to the next turn.}
    \label{fig:1}
\end{figure*}

\section{Problem Formulation}
\label{sec:formulation}

We formalize \textbf{SidConArena} as a finite-horizon Partially Observable Stochastic Game (POSG) defined by the tuple $\mathcal{G} = \langle \mathcal{N}, \mathcal{S}, \mathcal{A}, \mathcal{P}, H, V_{terminal},\mathcal{O}, Z \rangle$.
$\mathcal{N} = \{1, \dots, n\}$ is the set of asymmetric agents. $H$ represents the finite horizon. $\mathcal{S}$ is the global state space.
$\mathcal{A}$ is the joint action space. Reflecting the game's distinct phases, the action space for agent $i$ is composite: $\mathcal{A}_i = \mathcal{A}_{neg} \cup \mathcal{A}_{bid} \cup \mathcal{A}_{prod}$. We detail these phase-specific actions in the following subsections.
The transition dynamics $\mathcal{P}$ are mixed: \textit{Production} is a deterministic optimization process, while the \textit{Auction} phase introduces stochasticity and imperfect information (hidden bids).
Unlike standard RL settings with dense rewards, agents aim to maximize a terminal value function $V_{terminal}$ at step $H$. $\mathcal{O}_i$ denotes observation of $i$-th agent and $Z: \mathcal{S} \rightarrow \mathcal{O}$ is the observation function. Though from the rule agents

The game proceeds in alternating phases: a cooperative negotiation game, a deterministic local optimization, and a competitive auction game.

\subsection{Phase 1: Negotiation (Unstructured Cooperative Game)}
Agents start each turn with an \textit{Inherent Deficiency}: their internal production capacity is insufficient to fully utilize their converters. To resolve this, agents enter an unstructured bargaining phase to exchange resources $\mathbf{r}_{t,i} \in \mathbb{N}^d$.

Agents select negotiation actions $a_{neg} \in \mathcal{A}_{neg}$, consisting of natural language messages and binding trade vectors $\tau_{i \leftrightarrow j} = (\Delta \mathbf{r}_{i \to j}, \Delta \mathbf{r}_{j \to i})$.
This phase is modeled as a cooperative sub-game where trades occur if they increase the heuristic utility $U$ for both parties:
\begin{equation}
\label{eq1}
    U_i(s_i + \tau) > U_i(s_i) \land U_j(s_j - \tau) > U_j(s_j).
\end{equation}
Here, $U(s)$ is a heuristic valuation function (defined in Sec.~\ref{subsec:objective}) used to estimate the implicit value of resources before the terminal step.

\subsection{Phase 2: Production (Deterministic Optimization)}
Following negotiation, agents execute a local production plan. The state component includes a set of converters $\mathcal{F}_{t,i}$, where each $f \in \mathcal{F}_{t,i}$ is a function mapping input resources to outputs.
The agent selects an activation vector $\mathbf{x}_{i} \in \{0,1\}^{|\mathcal{F}_{t,i}|} \subset \mathcal{A}_{prod}$.
The transition is deterministic:
\begin{equation}
\label{eq2}
    \mathbf{r}_{t+1, i} = \mathbf{r}_{t,i} - \sum_{f} x_f \mathbf{r}_{current}(f) + \sum_{f} x_f \mathbf{r}_{out}(f).
\end{equation}
This phase is effectively a Knapsack-like combinatorial optimization problem, constrained by the inventory $\mathbf{r}_{t,i}$ accumulated during negotiation.

\subsection{Phase 3: The Confluence (Competitive Imperfect-Information Game)}
The allocation of permanent assets (Colonies and Technologies) is modeled as a \textbf{Simultaneous Multi-Track Priority Auction}. This is the primary source of imperfect information, as agents submit sealed bids using a liquid currency, \textit{Ships}.

Let $B_i$ be agent $i$'s budget of Ships. Agents submit a bid tuple $a_{bid} = (b_{i,\mathcal{C}}, b_{i,\mathcal{R}}) \in \mathcal{A}_{bid}$ subject to the budget constraint $\sum b \le B_i$. The allocation follows a \textit{Right-of-Choice} mechanism:
\begin{enumerate}
    \item \textbf{Ranking:} Agents are sorted descending by $b_{i,track}$. Tie-breaking introduces stochasticity based on asset holdings.
    \item \textbf{Selection:} In rank order, agent $i$ selects an item $m$ from available assets $M_{track}$, provided $b_{i,track} \ge p_{min}(m)$, where $p_{min}$ is the item's reserve price.
    \item \textbf{Payment:} The agent pays the full bid $b_{i,track}$. The overpayment $b_{i,track} - p_{min}(m)$ represents the premium paid for priority.
\end{enumerate}

\subsection{Objective and Valuation Functions}
\label{subsec:objective}
The ultimate goal $J$ is to maximize the total value of the agent's holdings at the terminal step $H$.
Let $\mathbf{w} \in \mathbb{R}^d$ be a ground-truth valuation vector assigning weights to all resources (including "Score" tokens, which typically have the highest weight $w_{score}$). The objective is:
\begin{equation}
\label{eq3}
    J(\pi_i) = \mathbb{E} \left[ V_{terminal}(s_{H,i}) \right] = \mathbb{E} \left[ \mathbf{w} \cdot \mathbf{r}_{H,i} \right].
\end{equation}
Note that there is no time-discounting for the terminal objective.
However, to make rational decisions during intermediate turns ($t < H$), agents must rely on a heuristic utility $U(s_{t,i})$ that estimates the \textit{Net Present Value} (NPV) of their state. This heuristic accounts for the immediate resource value plus the discounted future yield of their production engine $\mathcal{F}$:
\begin{equation}
\label{eq4}
    U(s_{t,i}) \approx \mathbf{w} \cdot \mathbf{r}_{t,i} + \sum_{k=t}^{H-1} \gamma^{k-t} \cdot \text{EstYield}(\mathcal{F}_{t,i}).
\end{equation}
This formulation clarifies the strategic tension: agents engage in auctions to acquire high-yield $\mathcal{F}$ (improving the second term of $U$) by spending resources (lowering the first term of $U$), ultimately converging to maximize $V_{terminal}$ at $t=H$.

\section{The SidConArena Framework}\label{sec::method}

Building on the finite-horizon POSG formulation in Section~\ref{sec:formulation}, \textit{SidConArena} instantiates the game as an executable multi-agent simulation framework. The framework is designed around a separation between agent-visible observation, LLM-based decision making, and rule-grounded environment execution. As summarized in \Cref{fig:1}, this design enables open-ended negotiation while preserving structured action validation and consistent state transitions.

\subsection{Environment State and Observation Construction}
% The environment acts as the authoritative state manager, enforcing the transition dynamics $\mathcal{P}$ across the three distinct game phases.
% The environment serves as the authoritative state manager, maintaining the global game state and transforming it into structured observations for each agent. This observation construction process corresponds to Stage~1 in \Cref{fig:1}. 

% \textbf{State Representation.}
% The environment maintains the global state $\mathcal{S}$, tracking the resource vector $\mathbf{r}_{t,i}$ for all agents. The vector dimension $d$ covers production resources, a liquid bidding currency, and terminal achievement indicators. To support heuristic valuation $U(s_{t,i})$ in \Cref{eq4}, the observation space also exposes private, public, market, and interaction information relevant to economic value estimation.
% The environment serves as the authoritative state manager, maintaining the global game state and transforming it into agent-specific structured observations. This process corresponds to Stage~1 in \Cref{fig:1}.

\textbf{State and observation.}
The global state $\mathcal{S}$ tracks each agent's resource vector $\mathbf{r}_{t,i}$, converter set $\mathcal{F}_{t,i}$, public board information, auction state, and active negotiation records. The resource vector dimension $d$ covers production resources, the liquid bidding currency, and terminal achievement indicators. At each decision point, the environment constructs an observation for agent $i$ by filtering the global state according to visibility and phase:
\[
\mathcal{O}_{t,i}
=
\{\mathcal{O}_{private}, \mathcal{O}_{public},
\mathcal{O}_{market}, \mathcal{O}_{interaction}\}.
\]
These four channels correspond to the four observation blocks in \Cref{fig:1}. They expose the agent's private inventory and production capabilities, public information about other agents and bulletin-board signals, phase-specific auction context, and active negotiation state. Together, they provide the information needed for local valuation, trade discovery, production planning, bidding, and item selection, while preventing agents from accessing hidden or unauthorized game state. Detailed field-level descriptions are provided in Appendix~\ref{app:observation_details}.

\subsection{Agent Architecture and Execution Pipeline}

The system architecture of \textit{SidConArena} is organized around a separation between stochastic agent reasoning and deterministic environment execution. As shown in \Cref{fig:1}, structured observations are routed by a phase-aware dispatcher to specialized reasoning modules, whose outputs are converted into executable function-calling actions through a neural-symbolic interface and subsequently applied by the environment-side execution engine. We additionally implement a web-based human interface that allows human players to participate in the same environment under the shared observation and action protocol; details are provided in Appendix~\ref{app:human_interface}.
% This design supports open-ended natural-language interaction while preserving strict control over action validity, resource constraints, and game-state transitions.

\textbf{Modular Brain and Context-Aware Dispatching.}
To manage the distinct sub-games defined in Section~\ref{sec:formulation}, we employ a hierarchical \texttt{Brain} architecture. Acting as a central dispatcher, the \texttt{Brain} routes the structured observation to specialized reasoning modules according to the current game phase. These callers handle turn planning, trading, production, bidding, and item selection, respectively. This modular design prunes irrelevant context and allows each LLM call to focus on the phase-specific decision problem. Detailed specifications of each caller are provided in Appendix~\ref{app:modules}.

\textbf{Neural-Symbolic Action Interface.}
To reconcile free-form natural-language reasoning with the rigid logic of the game engine, the framework enforces a structured action interface. Agents produce function calls rather than directly modifying the game state. This interface validates action syntax and preserves quantitative precision for trades, production plans, bids, and item selections.

% Agents may still generate free-form reasoning in a dedicated \texttt{reasoning} field before committing to an executable action.

\textbf{Asynchronous Event-Driven Execution.}
Standard turn-based multi-agent environments typically block execution until all agents respond. In contrast, the negotiation phase of \textit{SidConArena} requires high-frequency, simultaneous interaction. We therefore implement a non-blocking execution pipeline based on a client-server architecture and broadcast state updates through a publish-subscribe pattern. When an event occurs, such as receiving a trade proposal $\tau_{j \to i}$, the scheduler triggers an asynchronous task for the corresponding agent's \texttt{Brain}. This allows multiple agents to perceive, reason, and counter-offer concurrently without halting the global game loop. Submitted actions are then applied by the environment-side execution engine. We provide implementation details of these phase-specific execution engines in Appendix~\ref{app:env_details}.
% filepath: experiments_section.tex
\section{Experiments}
\label{sec:experiments}

\begin{figure*}[t]
    \centering
    \includegraphics[width=0.95\textwidth]{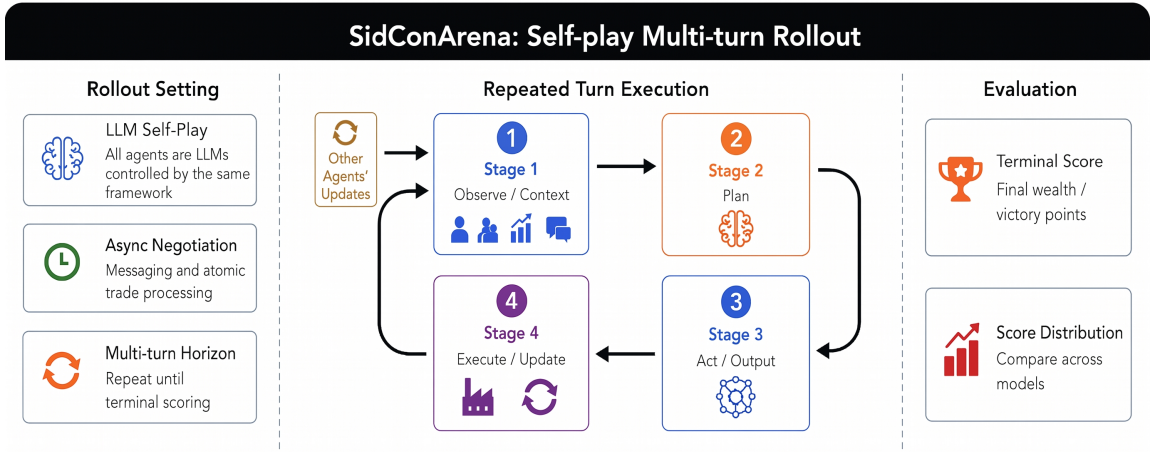}
    \caption{
    Self-play multi-turn rollout protocol in \textit{SidConArena}. 
    Each evaluation game instantiates LLM agents under the shared agent--environment pipeline in \Cref{fig:1} and repeats an observe--plan--act--execute/update cycle until terminal scoring. 
    Agents receive asynchronous updates from other players, emit phase-conditioned structured actions for negotiation, production, bidding, and asset selection, and are evaluated by terminal economic outcomes.
    }
    \label{fig:self_play_rollout}
\end{figure*}

We evaluate \textit{SidConArena} from two complementary perspectives. The first uses homogeneous self-play, where all seats in a game are controlled by agents instantiated from the same LLM backbone. The second uses heterogeneous tournaments, where different model backbones interact in the same economy and are ranked by an Elo-style rating computed from terminal outcomes. All agents use the same observation format, action schema, prompting interface, and environment rules. As illustrated in \Cref{fig:self_play_rollout}, each evaluation game repeatedly executes the shared observe--plan--act--execute/update loop until terminal scoring.

\paragraph{Models.}
Our model suite covers multiple model families, capacity regimes, and deployment paradigms, including compact agents such as GPT-4o-mini~\citep{openai2024gpt4ocard} and o3-mini, general-purpose chat models such as Qwen-Plus~\citep{yang2025qwen3technicalreport} and DeepSeek-V3~\citep{deepseekai2025deepseekv3technicalreport}, and recent high-capability systems such as Gemini-3-Flash-Preview, GPT-5~\citep {singh2026openaigpt5card}, and Claude-Opus-4.

\paragraph{Protocol and metrics.}
Each game follows the three-phase structure formalized in Section~\ref{sec:formulation}: agents negotiate and submit binding trades, activate production converters, and participate in sealed-bid auctions for colonies and technologies. The primary outcome is terminal economic performance, computed from the final value of each player's holdings. For heterogeneous tournaments, we additionally compute model-level Elo ratings from completed games. To diagnose behavioral failures, we analyze trajectory logs containing messages, structured actions, resource states, proposals, bids, allocations, and terminal scores. The detailed experimental settings and analysis procedure are provided in Appendix~\ref{app:analysis_procedure}.

\paragraph{Experimental Settings} We implement two experimental setups: Self-playing and Elo Tournaments. For the former, we run games where all species are controlled by the same model, aiming to obtain its baseline performance. For the latter, we randomly select different models playing each species in order to reduce performance biases caused by species differences. The Elo Tournament is run for a total of 20 games.
\section{Results}
\label{sec:results}

We now report the main empirical findings. Overall, LLM agents can operate within the \textit{SidConArena} environment, but their performance varies substantially across model backbones. Our analysis shows that strong economic performance depends on context-specific valuation, positive-sum coordination, strategic bargaining, and long-horizon investment planning.

\subsection{Homogeneous Rollouts Reveal Absolute Economic Performance}
\label{subsec:model_performance_gap}

\begin{figure}[h]
    \centering
    \includegraphics[width=\linewidth]{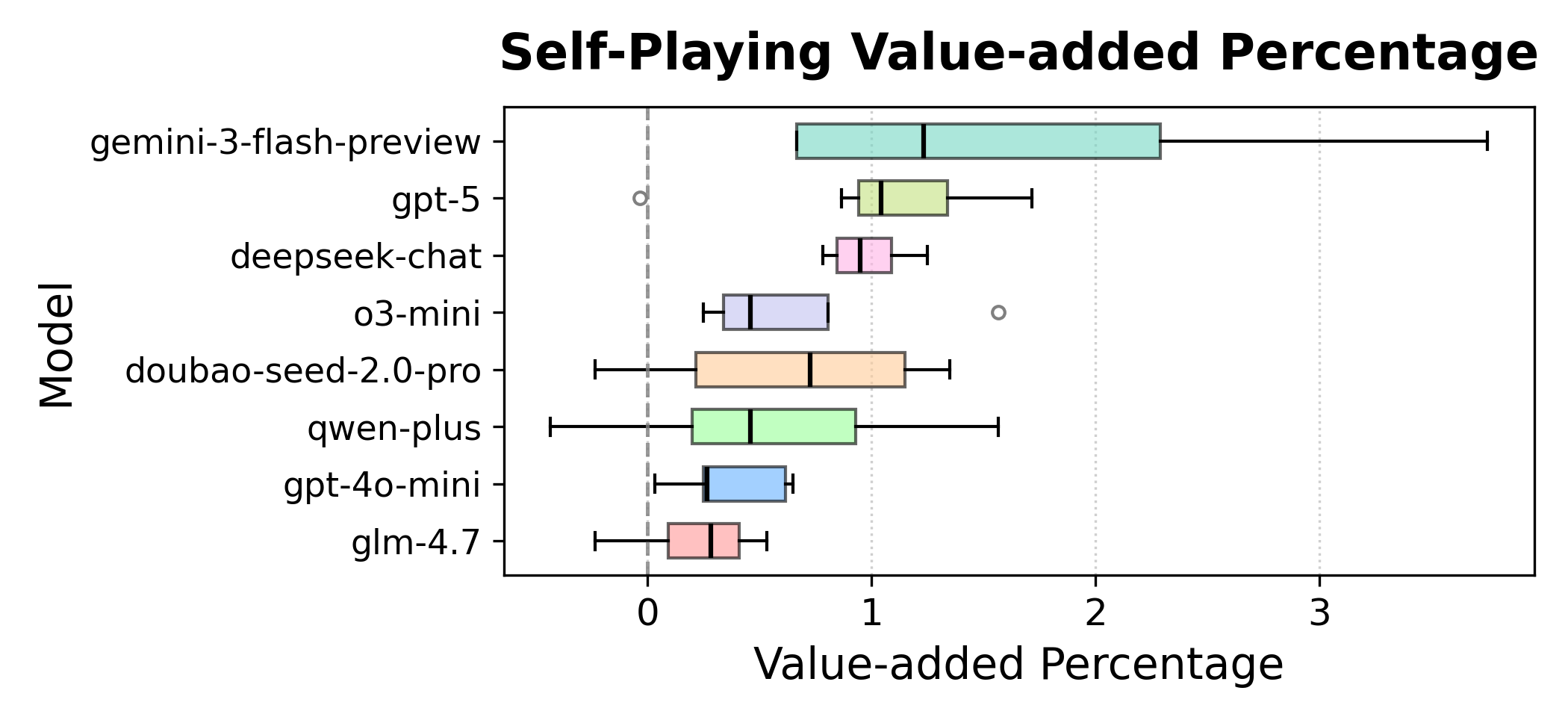}
    \caption{Distribution of self-play performance scores across different LLM backbones. Stronger frontier models tend to achieve higher scores, whereas lightweight or earlier-generation models obtain substantially lower economic returns.}
    \label{fig:value_added}
\end{figure}

Figure~\ref{fig:value_added} compares the distribution of self-play performance across different LLM backbones. The results show clear model-level variation: stronger frontier models generally obtain higher terminal scores, while smaller or earlier-generation models remain substantially less effective in the SidConArena economy.

Among the evaluated agents, high-capability models such as GPT-5 and Gemini-3-Flash-Preview achieve the strongest overall performance. These agents are better able to maintain coherent production plans, identify profitable trades, allocate resources across negotiation and production phases, and convert intermediate gains into terminal value. In contrast, lightweight models such as GPT-4o-mini obtain much lower scores, suggesting that general instruction following and surface-level task compliance  are insufficient for strong performance in open-ended positive-sum bargaining.

The weak performance of GLM-4.7 further suggests that SidConArena does not merely reproduce a chronological ordering of model generations. Instead, it introduces a distinct evaluation axis for open-ended economic agency, where agents must combine bargaining, local valuation, resource allocation, and delayed-return reasoning in a dynamic multi-agent economy. A model may be recent and generally capable, yet still perform poorly if it lacks this combination of bargaining, valuation, and long-horizon economic reasoning skills.

\begin{figure}[h]
    \centering
\includegraphics[width=\linewidth]{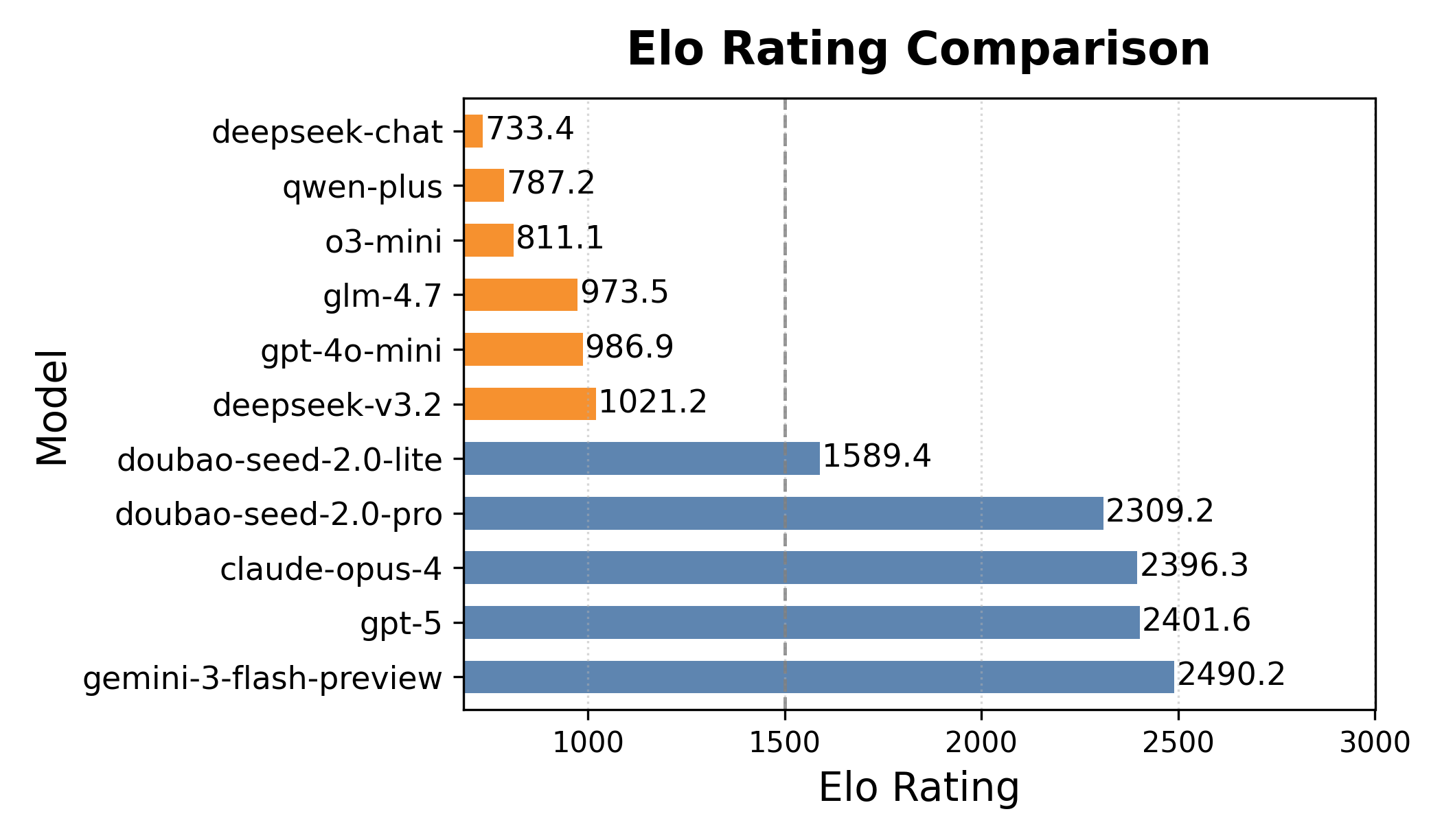}
    \caption{Elo of models. We apply 1500 as baseline elo, and the tournament is zero-sum.}
    \label{fig:elo_tournament}
\end{figure}

\subsection{Elo Tournament Reveals Relative Competitive Strength}
\label{subsec:elo_tournament}

While homogeneous rollouts measure absolute economic performance under controlled model assignments, they do not fully capture how agents perform when directly exposed to opponents with different reasoning patterns, bargaining styles, and valuation errors. We therefore conduct heterogeneous Elo tournaments, where different seats in the same game are controlled by agents instantiated from different model backbones.

Figure~\ref{fig:elo_tournament} reports the model-level Elo ratings computed offline from a batch of heterogeneous tournament games. Overall, the Elo ranking provides a complementary view to homogeneous rollout scores. Strong frontier models remain competitive in mixed-agent economies, indicating that their economic performance is not solely an artifact of interacting with similar agents. These models are better able to exploit favorable trades, contest scarce auction assets, and preserve terminal value even when facing opponents with different bargaining behavior.

% \begin{figure}[ht]
%     \centering
%     \includegraphics[width=0.9\linewidth]{fig/elo_tournament.png}
%     \caption{Offline Elo ratings from heterogeneous tournament games. Different seats in each game are assigned to different model backbones. Elo ratings summarize relative competitive strength across the collected tournament outcomes.}
%     \label{fig:elo_tournament}
% \end{figure}

The Elo results also reveal that absolute rollout performance and competitive strength are related but not identical. A model with high homogeneous scores may still lose relative advantage in heterogeneous games if it fails to adapt to unfamiliar opponents, overpays in contested auctions, or accepts trades that are locally plausible but strategically unfavorable. Conversely, some agents may achieve moderate homogeneous scores yet remain competitive in mixed settings if they bargain more assertively or allocate resources more robustly under opponent diversity. 
% This distinction highlights why \textit{SidConArena} benefits from both evaluation views: homogeneous rollouts measure a model's standalone economic agency, while heterogeneous Elo tournaments measure its strength in direct economic competition against other model agents.

\subsection{LLM Agents Can Execute the Game but Struggle with Local Valuation}
\label{subsec:rule_following_and_valuation}

Across collected self-play trajectories, modern LLM agents usually follow the high-level phase structure of \textit{SidConArena}. 
They can participate in negotiations, submit structured trade proposals, activate production converters, and make auction decisions through valid actions. 
This indicates that the neural-symbolic interface in Section~\ref{sec::method} is effective for grounding open-ended natural-language interaction in an executable game engine.

However, syntactic action validity does not imply economic competence. 
Agents can produce valid actions while still misapplying the local valuation rules specified by the environment. 
To inspect this failure mode, we conduct a targeted qualitative audit of Ship-related trades and proposals from recorded self-play trajectories. 
As shown in \Cref{tab:ship_bias_examples}, agents sometimes price \textit{Ships} above their local valuation baseline, suggesting that they may rely on semantic or strategic associations of \textit{Ships} as scarce bidding resources rather than consistently grounding decisions in the local economic model.

This valuation error indicates that agents may satisfy the structured action interface while failing to ground decisions in the task-specific marginal-value model. 
Because trades, production decisions, and auctions are coupled across turns, even small pricing errors can compound into weaker economic outcomes. 
We provide a more detailed discussion of this failure mode in Appendix~\ref{app:ship_bias}.

\begin{table*}[t]
\centering
\small
\setlength{\tabcolsep}{4pt}
\renewcommand{\arraystretch}{1.12}
\begin{tabular}{p{0.13\textwidth}p{0.30\textwidth}p{0.34\textwidth}p{0.16\textwidth}}
\hline
{\raggedright Agent (Pair)\par} &
{\raggedright Recorded exchange\par} &
{\raggedright Observed behavior\par} &
{\raggedright Valuation pattern\par} \\
\hline

% Evidence: logs/log_1779760708.747003.txt:1091530,1093087,1093096
{\raggedright gpt-4o-mini $\rightarrow$ deepseek-v3.2\par} &
{\raggedright
The agent trades $\mathrm{Industry}\mathord{\times}3$ for
$\mathrm{Ship}\mathord{\times}1$, although the local valuation table assigns
listed values of 3 and 1, respectively.\par} &
{\raggedright
The agent evaluates the deal as highly favorable because it obtains one Ship,
despite paying a three-unit resource bundle for a one-unit resource.\par} &
{\raggedright
Buyer-side Ship premium.\par} \\
\hline

% Evidence: logs/log_1779650958.281904.txt:235348,236037,236046
{\raggedright deepseek-v3.2 $\rightarrow$ claude-opus-4\par} &
{\raggedright
The agent trades $\mathrm{Ship}\mathord{\times}1$ for
$\mathrm{Culture}\mathord{\times}2$, although the listed values are 1 and 2,
respectively.\par} &
{\raggedright
The agent describes the offer as excellent value, even though the exchange
gives up a bidding resource at a two-to-one listed-value rate.\par} &
{\raggedright
Seller-side opportunity-cost mismatch.\par} \\
\hline

% Evidence: logs/log_1768799058.640042.txt:116,375,573
{\raggedright qwen-plus \par} &
{\raggedright
The agent proposes $\mathrm{Ship}\mathord{\times}1$ in exchange for
$\mathrm{Culture}\mathord{\times}4+\mathrm{Biotech}\mathord{\times}1$,
with listed values of 1 versus 5.5.\par} &
{\raggedright
The proposal is posted immediately after the prompt establishes the one-unit
Ship baseline, yet the agent frames the request as passing its fairness check.\par} &
{\raggedright
Extreme seller asking price.\par} \\
\hline

% Evidence: logs/log_1779629117.305784.txt:367895,367904,368122
{\raggedright claude-opus-4\par} &
{\raggedright
The agent offers $\mathrm{Food}\mathord{\times}2$ for
$\mathrm{Ship}\mathord{\times}1$, although the listed values are 2 and 1,
respectively.\par} &
{\raggedright
The agent advertises the quote as a fair-value trade while offering two Food
per Ship to acquire additional bidding currency.\par} &
{\raggedright
Buyer-side Ship premium.\par} \\
\hline
\end{tabular}
\caption{
Representative Ship-related valuation mismatches from recorded self-play trajectories.
The examples show accepted exchanges and posted proposals in which agents price \textit{Ships} above, or reason inconsistently with, their approximately one-unit local valuation baseline.
}
\label{tab:ship_bias_examples}
\end{table*}

\begin{figure*}[h]
    \centering
    \includegraphics[width=\textwidth]{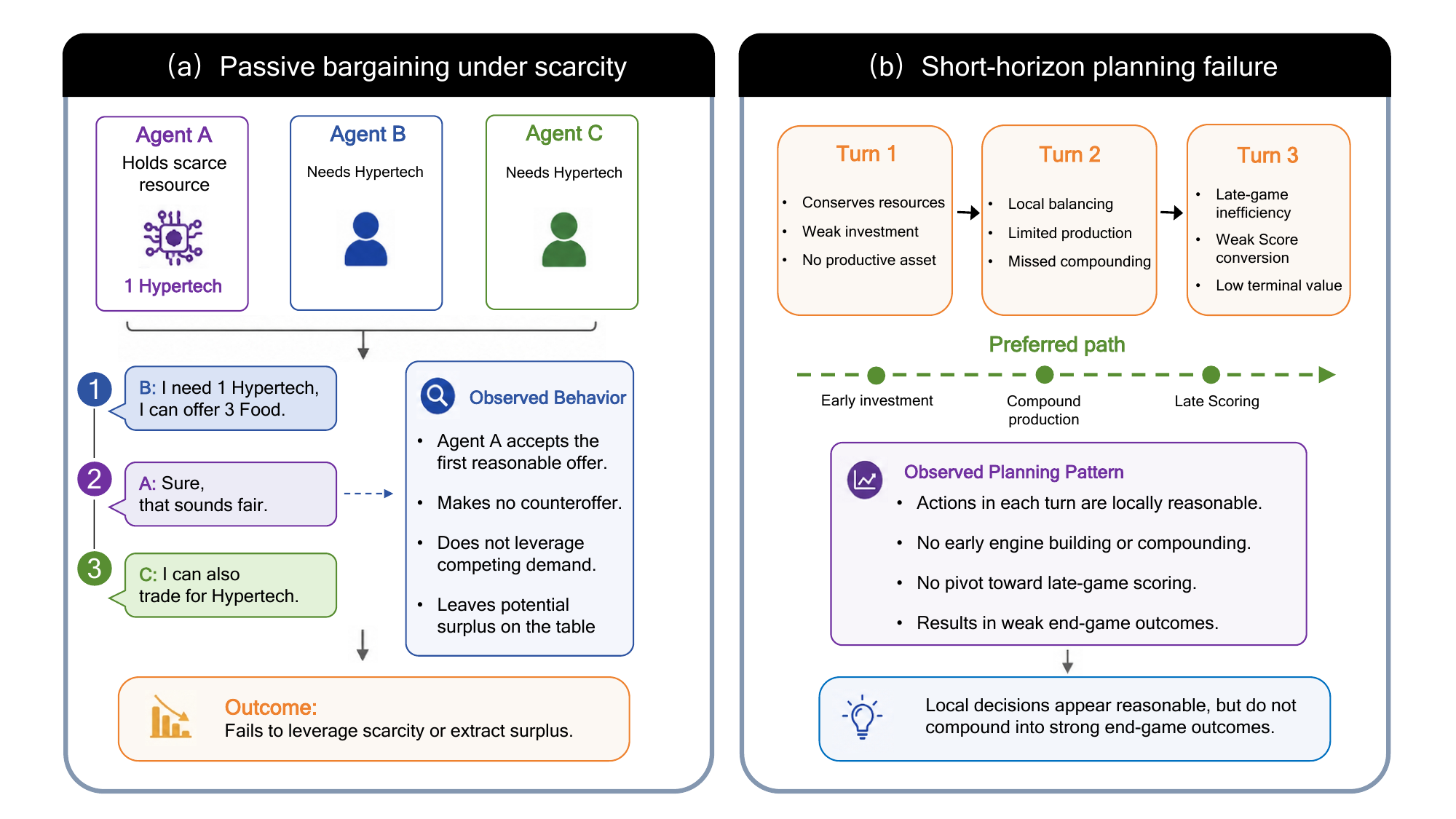}
    \caption{
    Representative qualitative patterns observed in \textit{SidConArena}. 
    (a) Passive bargaining under scarcity: even when holding a scarce resource demanded by multiple partners, the agent accepts the first reasonable offer without counteroffering or leveraging competing demand. 
    (b) Short-horizon planning failure: the agent's actions appear locally reasonable across turns, but the trajectory does not follow the preferred path from early investment to compound production and late-game scoring.
    }
    \label{fig:qualitative_failure_cases}
\end{figure*}

\subsection{Agents Exhibit Passive and Overly Polite Bargaining}
\label{subsec:passive_bargaining}

Beyond whether trades are accepted, successful play in \textit{SidConArena} requires agents to reason about bargaining power, temporary scarcity, and outside demand. Inspection of self-play trajectories suggests that LLM agents often remain cooperative in tone but passive in negotiation strategy. They tend to accept fair-looking offers, acknowledge other agents' needs, and avoid more strategic bargaining moves such as anchoring, counteroffering, or inducing competition among buyers.

\Cref{fig:qualitative_failure_cases}(a) illustrates this pattern. The focal agent holds a scarce Hypertech token, while multiple other agents express demand for it. However, after receiving an initial offer, the agent immediately accepts the proposal as reasonable rather than delaying commitment, soliciting alternative offers, or using the second buyer's demand to improve the exchange. This behavior leaves potential surplus on the table: the agent is socially cooperative, but does not strategically exploit scarcity.

This observation highlights a gap between politeness and bargaining competence. In open-ended positive-sum economies, agents must not only identify mutually acceptable trades, but also decide when to counteroffer, when to preserve optionality, and when to create competition among possible partners. Current LLM agents often satisfy the surface form of negotiation while failing to actively shape the market.

\subsection{Long-Horizon Investment Planning Remains a Major Failure Mode}
\label{subsec:long_horizon_planning}

A strong strategy in \textit{SidConArena} requires agents to coordinate decisions across multiple turns. Early turns should prioritize investment in productive capacity and resource engines; intermediate turns should use production to compound the value of earlier investments; and later turns should shift toward terminal score conversion. This temporal structure makes the task substantially harder than selecting locally beneficial trades or valid production actions in isolation.

\Cref{fig:qualitative_failure_cases}(b) shows a representative short-horizon planning pattern. The agent's decisions at each individual turn appear locally defensible: it conserves resources, balances inventory, and avoids obviously infeasible actions. However, the resulting trajectory fails to acquire productive assets early, misses opportunities for compounding production, and enters the late game with weak score-conversion capacity. The issue is therefore not merely invalid action generation, but a failure to connect local decisions into a coherent investment trajectory.

This pattern reflects a broader limitation of current LLM agents in delayed-return economic reasoning. Since \textit{SidConArena} rewards terminal value rather than immediate step-wise reward, agents must estimate how present trades, bids, and production choices affect future conversion opportunities. Agents that optimize only for local plausibility may appear competent at each phase while still producing weak end-game outcomes.

% \subsection{Summary}
% \label{subsec:results_summary}

% Our experiments reveal that LLM performance in \textit{SidConArena} depends not only on general model capability, but also on interactive economic reasoning. Frontier models such as GPT-5, Gemini-3-Flash-Preview, and Claude-Opus-4 show stronger economic agency, yet many agents still misprice resources, underexploit cooperative opportunities, and make myopic multi-stage decisions. These findings suggest that \textit{SidConArena} provides a compact and discriminative benchmark for evaluating economic agency in multi-agent LLMs.

% \input{sections/results}
% filepath: conclusion_future_work.tex
\section{Conclusion}
\label{sec:conclusion}

We introduced \textit{SidConArena}, a new benchmark framework for evaluating LLM agents in open-ended, positive-sum bargaining. SidConArena instantiates a mixed-motive economy in which agents must negotiate over complementary resources, activate production converters, and compete for scarce long-term assets through sealed-bid auctions. By combining structured observations, phase-aware agent dispatching, a neural-symbolic action interface, and asynchronous event-driven execution, the framework supports free-form multi-agent negotiation while maintaining validated, reproducible environment dynamics.

Our experiments show that SidConArena provides a discriminative evaluation setting for interactive economic agency. Stronger frontier models generally achieve better terminal outcomes in both homogeneous self-play and heterogeneous tournaments, indicating that the benchmark captures meaningful capability differences across LLM backbones. At the same time, trajectory-level analysis reveals that current agents remain far from robust economic decision makers. They often submit syntactically valid actions while misgrounding local resource values, accept socially reasonable offers without strategically exploiting scarcity, and fail to connect local decisions into coherent long-horizon investment plans.

These findings highlight a gap between surface-level agent competence and deeper economic reasoning. In SidConArena, success requires more than instruction following, role-playing, or isolated planning: agents must reason about context-specific value, bargaining power, opportunity cost, production synergies, and delayed terminal returns under partial observability. We hope SidConArena will serve as a useful testbed for developing and evaluating future LLM agents capable of more reliable positive-sum coordination, strategic negotiation, and long-horizon economic planning.

\section{Limitations}
\label{sec:limitations}

\paragraph{Scope of interaction evaluation.}
Although \textit{SidConArena} supports human participation through the unified interface, our quantitative evaluation focuses on LLM-agent rollouts, including homogeneous self-play and heterogeneous tournaments. 
We do not yet conduct a systematic study of mixed human--LLM games, where human players may actively exploit agent biases, adapt to agent bargaining styles, or reshape the negotiation dynamics. 
Extending the evaluation to controlled mixed-play settings is an important direction for measuring whether LLM agents can cooperate and compete robustly with human negotiators.

\paragraph{Simplified negotiation protocol.}
Our current implementation represents committed trades as binding structured proposals accompanied by natural-language messages. 
This design makes game execution reliable and enables precise trajectory analysis, but it does not fully capture the richer commitment structure of human negotiation. 
In particular, the environment does not yet model non-binding promises, conditional agreements, deferred compensation, reputation, or strategic betrayal. 
Future versions could introduce richer contract and trust mechanisms to evaluate whether agents can reason about credibility, reciprocity, and long-term social incentives.

\paragraph{Limited adaptation and training.}
Our experiments evaluate general-purpose LLM agents under a shared observation format, action schema, and prompting interface. 
We do not fine-tune models or train agents specifically for the \textit{SidConArena} economy. 
As a result, the reported results characterize the zero-shot or prompt-based economic agency of current LLMs, rather than the best performance achievable with environment-specific adaptation. 
Future work may investigate reinforcement learning, self-play training, or preference optimization to study whether agents can acquire stronger valuation, bargaining, and long-horizon planning strategies from repeated interaction.

\paragraph{Finite rollout scale and stochasticity.}
Because multi-agent bargaining games are expensive to run and analyze, our evaluation is necessarily based on a finite set of completed trajectories. 
The results reveal clear model-level differences and recurring behavioral failures, but individual games can still be affected by stochastic auction outcomes, asymmetric initial conditions, and interaction-specific negotiation paths. 
Larger-scale tournaments, repeated-seed evaluation, and more extensive human reference data would further improve the statistical robustness of model rankings and behavioral diagnostics.

\paragraph{Game-specific valuation and generality.}
Several observed failures, such as the overvaluation of \textit{Ships}, arise from the tension between pretrained semantic priors and the local valuation rules of this environment. 
While this makes \textit{SidConArena} useful for diagnosing context-specific economic reasoning, the benchmark is still grounded in a particular positive-sum board-game economy. 
Future extensions should test whether the same failure modes persist across other negotiation domains, resource systems, and market mechanisms, and whether explicit utility verification or tool-assisted valuation can reduce such biases.

% filepath: ethical_considerations.tex
\section{Ethical Considerations}
\label{sec:ethics}

% \subsection{Human Participation and Data Privacy}
% The human gameplay data referenced in Section~\ref{sec:results} (Benchmark Comparisons) and used for qualitative analysis was collected from volunteer participants. All participants were fully informed that their gameplay metrics, trading logs, and chat messages would be recorded for research purposes and provided explicit consent prior to participation.
% All Personally Identifiable Information were anonymized. No sensitive personal data was collected during the experiments.

% \subsection{Fair Use}
\textit{SidConArena} is a computational re-implementation of the game mechanics found in the board game \textit{Sidereal Confluence: Trading and Negotiation in the Elysian Quadrant}, designed by Brent Povis and published by WizKids.
Our environment reproduces the rules set and numerical values (converters, card stats), which are the factual mechanics of the game system. We do not distribute any copyrighted visual assets, artwork, or trademarked logos. This implementation is created strictly for non-commercial, academic research purposes to evaluate Artificial Intelligence in complex economic settings. 

Further statements are listed in Appendix~\ref{app:statements}.
% \section{Acknowledgements}\label{sec:ack}
% We sincerely thank Feifan Wang and Prof. Zhixuan Fang for their valuable discussions and insightful
% comments that help improve this work.

% We acknowledge the original designer and publisher for their creation of the underlying system that made this research possible.
% \newpage
% \section{Contribution}

% \begin{itemize}
%     \item Yeqi Feng: Front-end, Game Environment, \cref{sec:formulation}  
%     \item Yuxin Chen: Agent and Prompting, \cref{sec::method}  
% \end{itemize}
\bibliography{sidcon}

\appendix

\section{Statements}\label{app:statements}

\subsection{Potential Risks}

While our research focuses on foundational multi-agent evaluation within a simulated environment, we acknowledge the potential dual-use risks associated with developing and evaluating autonomous negotiation agents. If agents capable of open-ended, multi-player bargaining are applied to real-world scenarios—such as automated financial trading, algorithmic pricing, or automated procurement—without rigorous safeguards, they could inadvertently or intentionally engage in deceptive bargaining, collusive price-fixing, or the exploitation of human counterparts who lack equivalent AI assistance.

\subsection{Artifacts}

In this work, the primary scientific artifacts we utilize are large language models (LLMs) accessed via commercial APIs to drive the autonomous agents. We evaluate a diverse suite of models:

Azure OpenAI Models: We employ \texttt{gpt-4o-mini} (version \texttt{2024-07-18})~\citep{openai2024gpt4ocard}, \texttt{gpt-5} (version \texttt{2025-08-07})~\citep{singh2026openaigpt5card}, and \texttt{o3-mini} (version \texttt{2025-01-31}). These models are accessed via the Azure OpenAI API.

Volcengine (Ark) Models: We utilize \texttt{deepseek-v3-2-251201}~\cite{deepseekai2025deepseekv3technicalreport}, \texttt{doubao-seed-2-0-lite-260428}, \texttt{doubao-seed-2-0-pro-260215}, and \texttt{glm-4-7-251222}~\citep{5team2025glm45agenticreasoningcoding}. These are accessed via the Volcengine Ark API \footnote{\url{https://ark.cn-beijing.volces.com/api/v3}}.

Alibaba Cloud Models: We employ \texttt{qwen-plus}~\citep{yang2025qwen3technicalreport}, accessed via the Alibaba Cloud DashScope API \footnote{\url{https://dashscope.aliyuncs.com/compatible-mode/v1}}.

All model calls utilize default sampling parameters, except for setting \texttt{temperature=0} to reduce generation variance and ensure reproducible decision-making. We utilize these APIs strictly for research purposes to simulate multi-agent economic behaviors, and our usage fully complies with their respective access conditions and intended usage. Specifically, we adhere to the \emph{OpenAI Terms of Use}\footnote{\url{https://openai.com/policies/row-terms-of-use/}} and the \emph{Alibaba Cloud Platform Service Agreement}\footnote{\url{https://terms.alicdn.com/legal-agreement/terms/common_platform_service/20230728213935489/20230728213935489.html}}. We do not use or distribute the underlying weights of these models.

\subsection{LLMs Usage}
This work evaluates LLMs. Additionally, we use LLMs to improve the grammar and readability of this manuscript. The LLM was not used for any other scientific aspects of this work, and all intellectual content is solely the product of the authors.

\subsection{Budget}

In our simulations, each agent consumes an average of approximately 0.6 million tokens per complete game round.

\subsection{Packages}
We utilize LangChain\footnote{\url{https://github.com/langchain-ai/langchain.git}} 1.3.1 for establishing agents framework, with MIT license.

\section{Environment Implementation Details}

\subsection{Observation Construction Details}
\label{app:observation_details}

This appendix provides the field-level specification of the structured observation used in Stage~1 of \Cref{fig:1}. For each agent, the environment converts the global state into four observation channels according to information visibility and the current game phase.

\paragraph{Private endowment and production capabilities.}
The private channel $\mathcal{O}_{private}$ contains information needed for local valuation, production planning, and temporally consistent decision making. It includes the agent's current resource inventory $\mathbf{r}_{t,i}$, the converter set $\mathcal{F}_{t,i}$, species-specific information, and the turn-level strategic plan generated at the beginning of the current turn. Each converter entry specifies its input-output transformation logic, such as converting 3 \textit{Energy} into 1 \textit{Hypertech}, together with relevant meta-properties such as upgrade conditions, compatibility constraints, future investment requirements, and agent-specific production traits.

\paragraph{Public board and social signals.}
The public channel $\mathcal{O}_{public}$ exposes shared information that helps agents infer scarcity, trading intent, and bargaining opportunities. It includes observable resource counts and asset information of other agents, as well as bulletin-board messages. The bulletin board supports semi-structured cheap talk, including free-form declarations and explicit seeking/offering resource vectors, allowing agents to identify potential trading partners before binding negotiation.

\paragraph{Game phase context.}
The market channel $\mathcal{O}_{market}$ provides phase-specific information required for bidding and item selection. During the Bid Phase, agents observe the available auction tracks, including candidate items, floor prices in the bidding currency, and functional descriptions. This channel is only populated when the corresponding market information is relevant to the current phase.

\paragraph{Interaction information and state.}
The interaction channel $\mathcal{O}_{interaction}$ records the agent's active negotiation state. It includes incoming binding proposals $\tau_{j \to i}$ from other agents and the status of outgoing proposals initiated by the agent. This information allows the agent to evaluate offers, track pending commitments, avoid duplicate decisions, and maintain consistency across asynchronous negotiation events.

\subsection{Agent Modules}
\label{app:modules}
The \textbf{Valuation-Driven Neural-Symbolic Agent (VD-NSA)} architecture relies on a suite of specialized modules. The \texttt{Brain} architecture dispatches tasks to the following specialized modules:

\begin{description}
    \item[\texttt{TurnPlanCaller}]\ \\
    \textbf{Function:} Strategic Anchoring. \\
    \textbf{Trigger:} At the beginning of every turn $t$. \\
    \textbf{Role:} Generates the high-level "Long-term Plan" based on the current game state and species characteristics. This plan serves as a cognitive context for all subsequent actions in the turn, ensuring consistency between the economic goals and tactical execution.

    \item[\texttt{TradeCaller}]\ \\
    \textbf{Function:} Negotiation and Social Coordination. \\
    \textbf{Trigger:} During the Trading Phase (Asynchronous). \\
    \textbf{Role:} Handles natural language communication. It generates outgoing proposals to resolve resource deficiencies identified by the planner and evaluates incoming proposals from opponents to ensure Pareto improvements ($\Delta V > 0$).

    \item[\texttt{EconomyCaller}]\ \\
    \textbf{Function:} Resource Optimization. \\
    \textbf{Trigger:} During the Production Phase. \\
    \textbf{Role:} Solves the combinatorial challenge of converter activation. It determines the optimal sequence of \textit{Run} (resource conversion) to maximize the accumulation of needed resources, strictly adhering to the arithmetic constraints of the transition function.

    \item[\texttt{BidCaller}]\ \\
    \textbf{Function:} Bayesian Valuation. \\
    \textbf{Trigger:} During the Auction/Bid Phase. \\
    \textbf{Role:} Analyzes the public board state to estimate opponents' budgets and valuations. It outputs a bid vector $\mathbf{b}_i$ for Colony and Tech tracks, balancing the probability of winning against the risk of overpaying.

    \item[\texttt{PickCaller}]\ \\
    \textbf{Function:} Asset Selection. \\
    \textbf{Trigger:} During the Pick Phase (if a bid was successful). \\
    \textbf{Role:} Executes the "Right-of-Choice." Given the remaining items on the track and the agent's rank, it selects the specific Colony or Technology card that maximizes the marginal utility to the current inventory.

    \item[\texttt{DiscardColonyCaller}]\ \\
    \textbf{Function:} Constraint Management. \\
    \textbf{Trigger:} When the agent exceeds the colony ownership limit before the Production phase. \\
    \textbf{Role:} Performs a cost-benefit analysis to determine which asset to liquidate. It identifies the least efficient colony (e.g., one with poor input-output ratios or irrelevant production types) to discard, minimizing the loss of future production capacity.
\end{description}

\subsection{Three phases environment}
\label{app:env_details}

In this section, we detail the implementation logic for the three game phases within the SidConArena environment.

\textbf{Phase I: The Negotiation Ledger ($\mathcal{A}_{neg}$)}.\\
Corresponding to the cooperative sub-game, the environment provides an unstructured communication channel. It acts as a trusted third party to ensure atomic trade execution. When agents issue binding trade vectors $\tau_{i \leftrightarrow j} \in \mathcal{A}_{neg}$, the environment validates resource ownership and executes the transaction only upon mutual confirmation, strictly enforcing the cooperative constraint in \cref{eq1}.

\textbf{Phase II: The Production Engine ($\mathcal{A}_{prod}$)}. \\
This module implements the deterministic optimization logic. It accepts the binary activation vector $\mathbf{x}_i \in \mathcal{A}_{prod}$ from each agent. The engine validates constraints (ensuring input resources exist in $\mathbf{r}_{t,i}$) and applies the transition function in \cref{eq2}, converting inputs to outputs and updating the agent's inventory for the subsequent auction.

\textbf{Phase III: The Auction Mechanism ($\mathcal{A}_{bid}$)}.\\
To instantiate the imperfect-information competitive phase, the environment enforces a "sealed-bid" mechanism. It accepts bid tuples $(b_{i,\mathcal{C}}, b_{i,\mathcal{R}}) \in \mathcal{A}_{bid}$ simultaneously, ensuring no agent observes others' bids before submission. The system then executes the \textit{Right-of-Choice} priority algorithm, deducting the \textit{Ships} currency and distributing distinct assets ($\mathcal{C}, \mathcal{R}$) to determine the starting state for the next turn $\mathbf{r}_{t+1}$.

\subsection{Human Interface}
\label{app:human_interface}

\begin{figure}
    \centering
    \includegraphics[width=0.9\linewidth]{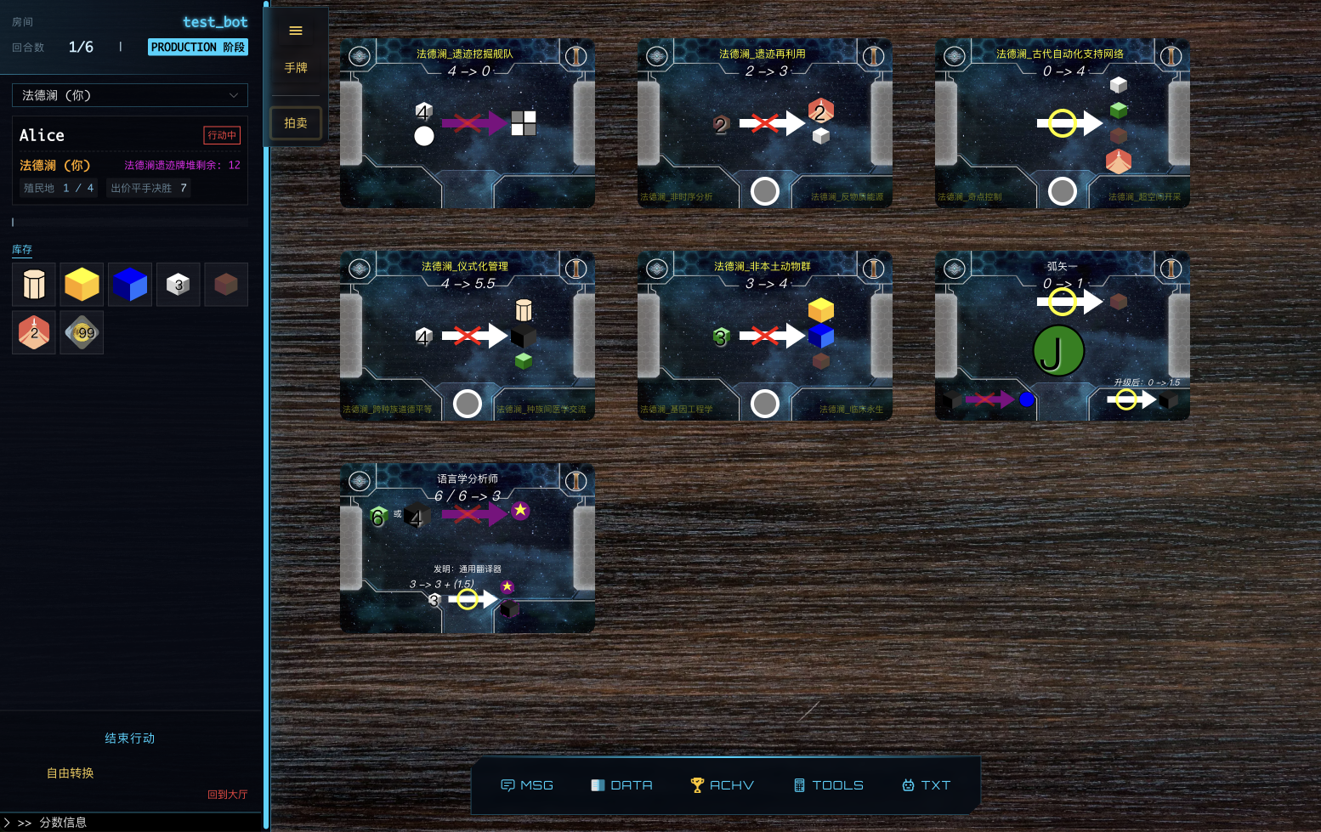}
    \caption{Web-based human interface for participating in \textit{SidConArena}.}
    \label{fig:human_interface}
\end{figure}

In addition to LLM self-play, \textit{SidConArena} provides a web-based interface for human participation. Human players connect to the same game server as LLM agents and receive the same structured observations, allowing them to participate under the shared environment state, action protocol, and game rules. This interface supports mixed-play sessions in which human players can negotiate, cooperate, or compete with LLM agents in real time. The frontend is implemented with Vue.js, as shown in \Cref{fig:human_interface}.

\section{Additional Analysis of Results}

\subsection{``Ship'' Valuation Bias}
\label{app:ship_bias}

Across collected self-play trajectories, we observe a recurring mismatch between the explicit resource valuation table and the reasoning used by LLM agents.
Although the environment specifies that a \textit{Ship} has an approximately one-unit local value, agents sometimes price Ships above this baseline or reason about them as intrinsically scarce strategic assets.

This bias appears to arise from the interaction between game mechanics and pretrained semantic priors.
Ships are used as bidding currency in the auction phase, so they can be strategically important in some contexts.
However, their immediate exchange value is still governed by the local valuation table.
Agents may therefore conflate the strategic role of Ships in auctions with their marginal exchange value during negotiation, leading to proposals or acceptances that deviate from the task-specific economic model.

\Cref{tab:ship_bias_examples} provides representative examples from recorded self-play trajectories.
In several cases, agents offer or accept multi-unit resource bundles for a single Ship while describing the exchange as fair or favorable.
These examples are not intended as a full-dataset estimate; rather, they illustrate a qualitative failure mode in which agents follow the structured action interface but misground local resource values.

This failure mode is particularly damaging in \textit{SidConArena} because bargaining, production, and auction decisions are coupled across turns.
A small pricing error in negotiation can affect resource availability for converter activation, bidding capacity in the auction phase, and ultimately terminal scoring.
The issue is therefore not merely whether an agent can produce valid structured actions, but whether it can internalize the environment-specific utility model instead of relying on pretrained lexical or strategic associations.

\section{Experimental Details and Analysis Procedure}
\label{app:analysis_procedure}

This section provides the experimental details omitted from the main text for space. All experiments are conducted in the \textit{SidConArena} environment described in Section~\ref{sec::method}. Each game follows the three-phase structure formalized in Section~\ref{sec:formulation}: agents first negotiate and submit binding trades, then activate production converters, and finally participate in a simultaneous sealed-bid auction for colonies and technologies. The environment acts as the authoritative state manager: it validates trade feasibility, enforces production constraints, resolves auctions, and records the full trajectory of messages, structured actions, resource states, bids, allocations, and terminal scores.

Agents receive observations through the structured interface illustrated in \Cref{fig:1}, which assembles private inventory, production capabilities, public board information, incoming and outgoing proposals, and phase-specific market information into a unified textual prompt. The same figure also summarizes the phase-aware dispatching and neural-symbolic action interface used during evaluation: model outputs are converted into structured function calls for trading, production, bidding, and item selection, and are then validated and executed by the environment engine. This design lets agents reason in natural language while ensuring that committed game actions are machine-checkable and executed under identical environment rules.

\paragraph{Evaluated agents.}
To obtain a broad and model-agnostic assessment of \textit{SidConArena}, we evaluate LLM agents spanning multiple model families, capacity regimes, and deployment paradigms. Our model suite includes compact and cost-efficient agents such as GPT-4o-mini, reasoning-oriented models such as o3-mini, strong general-purpose chat models such as Qwen-Plus and DeepSeek-V3, as well as more recent high-capability systems including DeepSeek-V4-Pro, Gemini-3-Flash-Preview, GPT-5, and Claude-Opus-4. All models are evaluated with the same environment interface, observation format, action schema, and game rules, without model-specific tools or privileged state access. Therefore, observed differences primarily reflect differences in model reasoning, negotiation, valuation, and long-horizon planning behavior.

\paragraph{Homogeneous model rollouts.}
Our first evaluation setting measures absolute economic performance under controlled multi-agent rollouts. For each model backbone, all seats in a game are controlled by agents instantiated from that same model, with identical prompting, observation, action interfaces, and environment rules. As shown in \Cref{fig:self_play_rollout}, each game repeatedly executes the shared agent--environment loop introduced in \Cref{fig:1}. This setting evaluates how well a model can sustain an economy when all players share the same reasoning capabilities, valuation tendencies, and bargaining style.

\paragraph{Heterogeneous Elo tournament.}
Our second evaluation setting measures relative competitive strength under the same rollout protocol, but with different model backbones assigned to different seats. In this setting, agents directly interact in a shared economy: they negotiate with heterogeneous opponents, compete for scarce auction assets, and adapt to different valuation and bargaining behaviors. After collecting a batch of heterogeneous tournament games, we compute model-level Elo ratings offline from the terminal outcomes of all participating agents. Compared with homogeneous rollouts, this tournament setting tests both whether an agent can translate its economic reasoning into strong outcomes and how competitive it is when directly matched against agents instantiated from other model backbones.

\paragraph{Terminal economic performance.}
Terminal score is the primary outcome measure for both evaluation settings. It is computed from the terminal value of each player's holdings after the final turn. Because \textit{SidConArena} rewards long-horizon engine building rather than isolated local gains, terminal score provides a compact measure of an agent's ability to trade, produce, invest, bid, and convert resources into end-game value.

\paragraph{Elo rating.}
To summarize relative competitive strength in heterogeneous tournaments, we compute an Elo-style rating from the collected tournament results. We estimate model-level ratings offline from a batch of completed games, using the terminal outcomes of all agents as the observed competition results.

\paragraph{Action validity and local valuation.}
To separate interface-level competence from economic competence, we record whether agents submit valid structured actions across trading, production, bidding, and item selection. We further inspect representative valuation errors by comparing selected trades and proposals against the game-specific resource valuation table, with a focus on Ship-related valuation mismatches.

\paragraph{Positive-sum cooperation.}
To quantify cooperative economic behavior, we analyze accepted and rejected trade proposals from the negotiation phase. We measure how often agents propose mutually beneficial trades, accept Pareto-improving offers, make counteroffers instead of ending negotiation, and trade with multiple partners. We also examine production utilization after negotiation, since underused converters indicate that agents failed to acquire the resources needed to unlock their own production engines.

\paragraph{Qualitative trajectory analysis.}
Beyond aggregate scores and production-utilization statistics, we qualitatively inspect trajectory logs to characterize strategic behaviors that are difficult to reduce to a single scalar metric. For bargaining, we examine whether agents passively accept fair-looking offers or actively counteroffer, preserve optionality, and leverage temporary scarcity. For long-horizon planning, we inspect whether turn-level decisions form a coherent trajectory from early investment to compounding production and late-game score conversion.

\paragraph{Analysis procedure.}
For quantitative analysis, we aggregate terminal scores by model backbone in homogeneous self-play and compute Elo ratings from heterogeneous tournament games. For diagnostic analysis, we parse the recorded trajectories produced by the pipeline in \Cref{fig:1}. The same rollout records therefore support both outcome-level evaluation and behavior-level diagnosis. This combined protocol is important because many failures in \textit{SidConArena} are strategic rather than syntactic: an agent may submit valid actions while still mispricing resources, refusing beneficial trades, bargaining passively, overpaying in auctions, or failing to exploit compounding production returns.

\end{document}